\newcommand{\ket}[1]{\mbox{$| {#1} \rangle$}}
\newcommand{\bra}[1]{\mbox{$\langle {#1} |$}}

\newcommand{\wo}{\mbox{$\omega_0$}}
\newcommand{\wk}{\mbox{$\omega_k$}}
\newcommand{\ak}{\mbox{$a_k^{}$}}
\newcommand{\adk}{\mbox{$a_k^\dagger$}}

\documentclass[prb,showkeys]{revtex4}

\begin{document}

\title{Entanglement induced by tailored environments}
\author{Markus A. Cirone, G. Massimo Palma}
\affiliation{NEST - CNR (INFM)  \& Dipartimento di Scienze Fisiche ed Astronomiche,
Universit\`{a} di Palermo, via Archirafi 36, I-90123 Palermo, Italy\\
{\tt massimo.palma@fisica.unipa.it}}
\date{\today}

\begin{abstract}
We analyze a system consisting of two spatially separated quantum
objects, here modeled as two pseudo-spins, coupled with a
mesoscopic  environment modeled as a bosonic bath. We show that by
engineering either the dispersion of the spin boson coupling or
the environment dimensionality - or both - one can in principle
tailor the spatial dependence of the induced entanglement on the
spatial separation between the two spins. In particular we
consider one, two and three dimensional reservoirs and we find
that while for a two or three dimensional reservoir the induced
entanglement shows an inverse power law dependence on the spin
separation, the induced entanglement becomes separation
independent for a one dimensional reservoir.

\end{abstract}

\maketitle

\section*{Introduction}

Among the various schemes that have been recently analyzed to generate entanglement between quantum subsystems,
 particular attention has been devoted to mechanisms based on the the interaction with  a so called entanglement mediator, i.e. a third system which interacts locally with the subsystems one wants to entangle (see e.g. \cite{Falci, Blais, Wallraff}).  Interestingly enough the role of entanglement mediator can be played by a large quantum reservoir. For example it has been shown that two atoms or two quantum dots interacting with the quantized electromagnetic field become steadily entangled even at finite temperature \cite{Braun}. At the same time  it has been  recently shown how cold atoms can be used to engineer, with a high degree of flexibility, systems described by spin boson Hamiltonians \cite{Zoller}. This opens the possibility to implement  new schemes in which  the desired entanglement between two microscopic quantum objects can be controlled by a suitable  design of a mesoscopic reservoir.
 In the present paper we will briefly analyze how, by engineering either the dispersion of the spin coupling or the dimensionality of the reservoir (or both) one can control the spatial dependence of the entanglement induced by a bosonic bath on two spatially separated quantum objects.
 
 \section*{The model}
To illustrate the idea consider a system consisting of two two-level systems, labeled $\alpha,\beta$ located at positions ${\bf r}_{\alpha}, {\bf r}_{\beta}$ respectively and interacting with a bosonic scalar quantum field.  We describe such a  system by the following   spin boson Hamiltonian
\begin{equation}
H=H_0+H_i
\end{equation}
with
\begin{eqnarray}
H_0 & = & \frac{\wo}{2} \left( \sigma_z^{\alpha}+\sigma_z^{\beta} \right) + \sum_{k}\wk \adk \ak \\
H_i& = & \sum_{i=\alpha\beta}\sum_{k} \left[ \left( \lambda_k({\bf r}_i) \ak
+\lambda_k^*({\bf r}_i)\adk\right) \left( \sigma_+^{i}+\sigma_-^{i}
\right) \right] \label{ham}
\end{eqnarray}
where  $\wo$ is the energy difference between ground   $\ket{g}$ and excited state $\ket{e}$ (we assume $\hbar=1$),
$\sigma_{z} = \ket{e}\bra{e} - \ket{g}\bra{g}$ , $\sigma_{+}=\ket{e}\bra{g}$ and $\sigma_{-}=\ket{g}\bra{e}$ are pseudospin operators and  $\ak$
and $\adk$ denote bosonic annihilation and creation operators respectively, of environment quanta of energy $\wk$.
 The  $\lambda_k({\bf r}_i)$ are position and frequency dependent coupling constants between the  atom $j$ and field mode $k$ and contain all the information on the spatial dependence of the quantized field modes. We will leave their specific functional form unspecified as long as possible in order to simplify the calculations and to obtain expressions which are as general as possible.
Note incidentally that the interaction Hamiltonian $H_i$  differs from the
one  examined in \cite{Braun}.

Straightforward time independent perturbation theory \cite{saku}, leads to the following expressions for the the perturbed (dressed)
ground state $\ket{\Psi}$ of the overall system system and for the interaction energy
$\Delta$ 

\begin{eqnarray}
\ket{\Psi} & = & c_{gg}\ket{g_{\alpha}g_{\beta}}\ket{0_k}+\frac{Q}{E_0-H_0}\left(
H_i-\Delta \right)\ket{\Psi} \label{dress}\\
\Delta & = & \bra{g_{\alpha}g_{\beta}} H_i \ket{\Psi}
\end{eqnarray}
where $\ket{g_{\alpha}g_{\beta}}\ket{0_k}$ is the unperturbed ground state with energy $E_0=- \omega_0$
and $Q={\cal I}-\ket{g_{\alpha}g_{\beta}}\bra{g_{\alpha}g_{\beta}}$ is a projector acting on the atomic degrees of freedom. The state equation (\ref{dress}) is in closed form and
can be expanded at the desired order of approximation. The
coefficient $c_{gg}$ is chosen in order to normalize $\ket{\Psi}$.

The normalized, "dressed" ground state of the system can be written
at second order of approximation in the following compact form

\begin{eqnarray}
\ket{\Psi^{(2)}} & = & c_{gg} \ket{g_{\alpha}g_{\beta}}\ket{0_k} +c_{ee}\ket{e_{\alpha}e_{\beta}} \ket{0_k}
+ \sum_{k}c_{eg,k} \ket{e_{\alpha}g_{\beta}}\ket{1_k}
+\sum_{k}c_{ge,k} \ket{g_{\alpha}e_{\beta}}\ket{1_k} \nonumber \\
& & + \frac{1}{2}\sum_{kk'}c_{gg,kk'}\ket{g_{\alpha}g_{\beta}}\ket{1_k 1_{k'}}
+ \frac{1}{2}\sum_{kk'}c_{ee,kk'}\ket{e_{\alpha}e_{\beta}}\ket{1_k 1_{k'}}
\label{psi}
\end{eqnarray}

\section*{The ground state entanglement}

To quantify the entanglement induced by the reservoir between the
two pseudospins we will use the two-tangle, i.e. the square of the
concurrence. We remind that given the density operator $\rho_{\alpha\beta}$
of a bipartite system of two qubits, the 2- tangle $\tau_{\alpha |\beta}$ is
defined as
\begin{equation}
\tau_{\alpha |\beta}(\rho)=[\max\left\{0,\xi_{1}-\xi_{2}-{\xi_{3}}-{\xi_{4}}\right\}]^2,
\end{equation}
where $\left\{\xi_{i}\right\}$ ($i=1,..,4$) are the square roots of the eigenvalues (in non-increasing order) of the
non-Hermitian operator $\bar{\rho}=\rho(\sigma_{y}\otimes\sigma_{y})\rho^{*}(\sigma_{y}\otimes\sigma_{y})$, $\sigma_{y}$ is the
$y$-Pauli operator and $\rho^{*}$ is the complex conjugate of $\rho$, in the eigenbasis of $\sigma_z  \otimes\sigma_z$ operator.
For states of the form like (\ref{psi}), the 2-tangle  takes the following simple form \cite{us}:

\begin{eqnarray}
\tau_{\alpha |\beta} & = & 4 \; |
\bra{e_{\alpha}e_{\beta}0_k}{\Psi^{(2)}}|^2 = 4 \; |
\bra{e_{\alpha}e_{\beta}0_k}\frac{Q}{E_0-H_0}H_i\frac{Q}{E_0-H_0}H_i\ket{g_{\alpha}g_{\beta}0_k}|^2
\nonumber \\
& = & \frac{4}{\omega_0^2} \left| \sum_k \frac{\lambda_k({\bf
r}_{\alpha})    \lambda^*_k({\bf r}_{\beta})}{\omega_0+\omega_k}\right|^2 \label{tangle}
\end{eqnarray}

In order to extract information on the dependence of the induced
entanglement on the spatial separation between the two spins we
must now specify the functional form of the coupling constants
$\lambda_k({\bf r}_i)$. As far as the mode structure is concerned,
we assume the usual plane-wave expansion

\begin{equation}
\lambda_k({\bf r}_i) \rightarrow \lambda_{\bf k}({\bf r}_i) =\epsilon_{\bf
k}\frac{e^{i {\bf k} \cdot {\bf r}_i}}{\sqrt{V}}
\end{equation}
The coupling factor $\epsilon_{\bf k}$ contains spin parameters
which we do not need to specify for our purposes. We will simply assume that it depends on the field mode frequency
spectrum via a power law, i.e.

\begin{equation}
\epsilon_{\bf k} \propto \omega_{\bf k}^\nu
\end{equation}
where ${\bf k}$ is the momentum of the oscillators. When
$V\rightarrow \infty$ the sum over momenta ${\bf k}$ can be
replaced by an integral

\begin{eqnarray}
\sum_k \frac{\lambda_k({\bf r}_{\alpha})\lambda^*_k({\bf r}_{\beta})}{\omega_0+\omega_k} &
\rightarrow & \frac{1}{(2\pi)^3} \int \!\!\! d{\bf
k}\frac{\epsilon_{\bf k}^A\epsilon_{\bf k}^B e^{i{\bf k} \cdot
{\bf r}}}{\omega_0+\omega_{\bf k}} \propto \int \!\!\! dk \;
k^{d-1}\frac{k^{2\nu} f(kr) }{k_0+k} \label{sumgg}
\end{eqnarray}
where ${\bf r}={\bf r}_A-{\bf r}_B$, $d$ is the spatial dimension
and a linear dispersion relation between $\omega_{\bf k}$ and
$k=|{\bf k}|$ has been assumed. The function $f(kr)$ comes from
the angular integration $\int d\Omega_{\bf k}$ and depends both on
the spatial dimension $d$ and on the dispersion law of the coupling. In
the atom-radiation interaction the coupling has vectorial
character and an $r^{-6}$ scaling of the tangle is found \cite{craig}. Here for the sake of
simplicity we assume a scalar coupling. We examine now in some
detail the dependence of energy and tangle on the spatial
dimension $d$ and on the value of $\nu$.

\subsection*{$d=3$}
In three dimensions the angular integration in (\ref{sumgg}) gives
\begin{eqnarray}
f(kr) & = & \int d\Omega_{\bf k} e^{i{\bf k} \cdot {\bf r}} = \int
\!\! d\phi \int \!\! d\theta \sin \theta e^{ik r \cos \theta} = 4
\pi \frac{\sin kr}{kr}, \;\;\;\;\; \label{angular3d}
\end{eqnarray}
Assuming $\nu=1/2$, as in the case of the coupling with the electromagnetic field, we obtain
\begin{eqnarray}
\sum_j \frac{g_j({\bf r}_A)g^*_j({\bf r}_B)}{\omega_0+\omega_j} &
\propto & \frac{1}{r}\int \!\!\! dk \; k^{2}\frac{\sin kr }{k_0+k}
\label{sum3d}
\end{eqnarray}
The integral present in (\ref{sum3d}) formally diverges. The
introduction of a cutoff momentum $k_c$ (always present in real
physical systems) gives
\begin{eqnarray}
\int_0^{k_c} dk \frac{k^2\sin kr}{k+k_0} & = & \frac{1}{r^2}  \left\{ \left( k_0-k_c \right)r
\cos k_cr + \sin k_cr +\right. \\
&+& k_0r \left[ -1  \right.  +k_0r \sin k_0r \left( \mbox{Ci}(k_0r) - \mbox{Ci}(k_0r+k_cr) \right)
\left. \left. +k_0r \cos k_0r \left( -\mbox{Si}(k_0r) + \mbox{Si}(k_0r+k_cr) \right)\right] \right\} \nonumber
\end{eqnarray}
Neglecting the unphysical fast oscillating terms  $\cos
k_c r$ and $\sin k_c r$ and and keeping the leading terms in $k_0
r \ll 1$, the tangle scales as $r^{-4}$ with the distance.
Clearly, different couplings lead to different scaling laws. For
instance, for flat, i.e. frequency independent coupling , $\nu=0$ and
the tangle scales as $r^{-2}$.

\subsection*{$d=2$}
In two dimensions we have
\begin{eqnarray}
f(kr) & = & \int d\Omega_{\bf k} e^{i{\bf k} \cdot {\bf r}} = \int
\!\! d\phi e^{ik r \cos \phi} = 2 \pi J_0(kr), \;\;\;\;\;
\label{angular2d}
\end{eqnarray}
where $J_0$ denotes the Bessel function of order $0$. Thus
\begin{eqnarray}
\sum_j \frac{g_j({\bf r}_A)g^*_j({\bf r}_B)}{\omega_0+\omega_j} &
\propto & \int \!\!\! dk \; k^{1+2\nu}\frac{J_0(kr) }{k_0+k}
\label{sum2d}
\end{eqnarray}
Numerical calculations show that this quantity scales
approximately as $r^{-1}$ for $\nu=0$ and $\nu=1/2$, and as
$r^{-3}$ for $\nu=1$.

\subsection*{$d=1$}
For a one-dimensional environment we get

\begin{eqnarray}
\sum_j \frac{g_j({\bf r}_A)g^*_j({\bf r}_B)}{\omega_0+\omega_j} &
\propto & \int_{-\infty}^\infty \!\!\! dk \;
k^{2\nu}\frac{e^{ikr} }{k_0+k}
\end{eqnarray}

Assuming $2\nu=n$, where $n$ is an integer, we get

\begin{eqnarray}
\sum_j \frac{g_j({\bf r}_A)g^*_j({\bf r}_B)}{\omega_0+\omega_j} &
\propto & \int_{-\infty}^\infty \!\!\! dk \; k^n\frac{e^{ikr}
}{k_0+k} = \int_{0}^\infty \!\!\! dk \; k^n\frac{e^{ikr} }{k_0+k}-\left( -1\right)^n
\int_{0}^\infty \!\!\! dk \; k^n\frac{e^{-ikr} }{k-k_0} \nonumber \\
& & = \frac{1}{i^n}\frac{\partial}{\partial r^n} \left[ \int_{0}^\infty \!\!\! dk \; k^n\frac{e^{ikr} }{k_0+k}-
\int_{0}^\infty \!\!\! dk \; k^n\frac{e^{-ikr} }{k-k_0} \right]=\left( -1 \right)^n i\pi k_0^n
e^{-ik_0r}
\end{eqnarray}
i.e. the tangle turns out to be independent of the spin separation $r$.

\section*{conclusions}

In summary we have shown that the interaction of two spatially
separated pseudospins  with a common environment leads to an
entanglement whose scaling on the spin spatial separation  depends on the dimensionality of the
environment and on  how the coupling constants scale with respect to the 
frequency of the reservoir modes.
This result holds for the class of physical systems
described by a spin-boson  Hamiltonian of the form (\ref{ham}). In
particular, for a  three and two dimensional environment we have
found a dependence which scales as $r^{-n}$, with different values
of $n$ depending on the dispersion of the coupling constants. On
the contrary, in one dimension the tangle does not depend on the
distance between the two spins.

Note that our analysis has been carried on by a perturbative expansion of the ground "dressed" state of the overall atoms field system. In other words we have studied the equilibrium state of the overall system. Although a full time dependent analysis of the atomic entanglement during the process of approach to equilibrium is out of the scope of the present articles we would like to point out the difference between our approach and some of the existing literature.  Some aspects of the time-dependent entanglement which builds up in the collective decay of two subsystems interacting with a common bath has been analyzed in \cite{benatti, tanas}. In particular they both study the time dependent as well as the asymptotic entanglement relation with the collective damping rate. In both studies however, consistently with the Wigner Veisskopf approach to the decay process, the ground state is not dressed. Here instead we have focussed our attention on the role of the bath as an entanglement mediator in the ground state. Of course, in the presence of collective interactions with a reservoir, there may exist subradiant states which are effectively decoupled from the environment. The entanglement which would characterize such state is out of the scope of this manuscript and cannot be analyzed with the techniques used above.

\acknowledgments
We acknowledge support from the  PRIN 2006 "Quantum noise in mesoscopic systems"  
and from EUROTECH S.p.A.

\end{document}